# A Fog Computing Based Architecture for IoT Services and Applications Development

Yousef Abuseta

*Computer Science Department, Faculty of Science, Zintan University*
*Zintan, Libya*

***Abstract** — IoT paradigm exploits the Cloud Computing platform to extend its scope and service provisioning capabilities. However, due to the location of the underlying IoT devices which is far away from the cloud, some services cannot tolerate the possible latency resulted from this issue. To overcome the latency consequences that might affect the functionality of IoT services and applications, the Fog Computing has been proposed.*

*Fog Computing paradigm utilizes local computing resources locating at the network edge instead of those residing at the cloud for processing data collected from sensors linked to physical devices in an IoT platform. The major benefits of such paradigm include low latency, real-time decision making and an optimal utilization of available bandwidth. In this paper, we offer a review of the Fog computing paradigm and in particular its impact on the IoT application development process. We also propose an architecture for Fog Computing based IoT services and applications.*

## I. INTRODUCTION

Fog Computing (FC), first introduced by Cisco, extends Cloud Computing by deploying locally computing and processing facilities into the edge of the network. This yields many benefits including location-awareness, low latency, and on time analytics for mission critical applications [1][2]. The Fog Computing nodes, which represent the resources and infrastructure of FC, are located between the physical devices at the network edge and the cloud.

The idea is to allow devices to talk directly to each other without the need to send data all the way to the cloud, enabling real-time decisions to be made and also shielding the IoT application from transmitting massive amount of data to the cloud. The FC objective is also to connect all devices to the cloud with open communication standards [3]. We believe that most IoT services and application are of real-time nature and thus require performing data processing and decision making in a timely manner. We also believe that IoT applications are dynamic and constantly changing at runtime in terms of the system requirements and the availability of the devices and their services. The engineering of such systems is usually carried out by performing some activities within a closed control loop from the area of control theory. Such activities are referred to as *collect*, *analyze*, *decide* and *act* as in [4] or *monitor, analyze, plan* and *execute* as in the IBM architecture blueprint [5].

In this paper, we investigate the characteristics of the Fog Computing paradigm and particularly its impact on architecting and designing IoT applications. We also propose an architecture for IoT applications residing at the Fog Computing platform.

The remainder of the paper is organized as follows. Section II reviews some background issues related to our proposed architecture. Section III introduces our proposed architecture for IoT Applications. In section IV, an evaluation case study is presented to illustrate the applicability of the proposed architecture. Section V reviews some of the previous works that have been conducted so far. The paper is concluded in section VI with some suggestions for further research.

## II. BACKGROUND

### A. Fog Computing Architecture

To better understand the importance of Fog Computing paradigm and its role in facilitating the provisioning of IoT services in a timely manner, this section is dedicated to introduce a high level architecture of this platform highlighting its fundamental components and characteristics. In a definition by [6], Fog Computing is " *a wireless distributed computing platform in which complex latency sensitivity tasks can be processed via a group of sharing resources at IoT gateway level in a locality* ". In another definition by [7], Fog Computing is " a *horizontal architecture on system-level that distributes computation, storage, control and networking capabilities closer to users along a cloud-to-device continuum*".

These two definitions reveal some fundamental issues related to the mechanism and architecture of Fog Computing model. Firstly, the computation and storage capabilities are distributed over a number of IoT devices that are located proximate to the device layer. Secondly, the emergence of FC was primarily driven by the desperate need of reducing (or optimizing) the processing and analysis time of collected data taken place in the cloud platform. This results in the realization of a real time response and decision making process. The computation, storage

Thirdly, the fog computing model resides between the device layer and the cloud. Fig.1 depicts a high level architecture of the Fog Computing paradigm. It shows how a set of disparate IoT devices can employ the fog computing to communicate with the cloud platform.

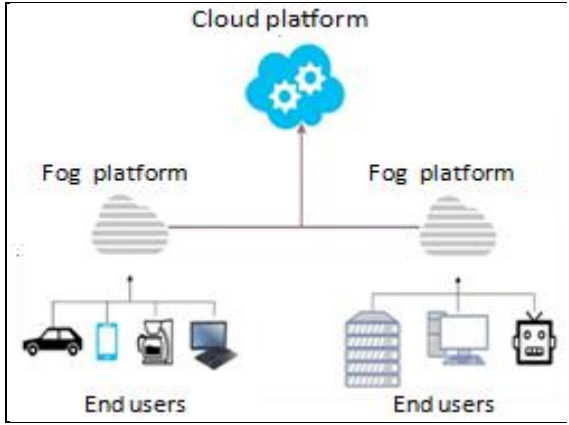

**Fig 1: A high level architecture of Fog Computing model [9].**

### B. Benefits of Fog Computing

As pointed out earlier in this paper, Fog Computing is an extension to the traditional cloud based platform since some functions are better performed in the cloud whereas others are obviously more advantageous to be carried out in the Fog Computing platform. Here are some situations where one paradigm is more suitable than the other:

- Time sensitive applications are better hosted and executed at the Fog computing platform. In such applications, data generated by sensors are stored, processed and analyzed in a timely manner and consequently decision making and any possible corrective actions (via actuators) are performed at the right time.
- The device management process at the Fog computing brings benefits to both the application under development as well as the cloud platform. Since the device management is done locally, the cloud is relieved from keeping track of a huge number of physical devices involved in the IoT paradigm.
- Big data, generated by a great number of smart devices, analytics tools are better hosted on the cloud platform since these tools require powerful computation and storage capabilities to run software such as machine learning algorithms.

### C. Managed Element and Autonomic Manager

In this section, we introduce some important concepts related to architecting fog computing based IoT applications. This architecture is inspired by the IBM feedback control loop introduced to engineer the and networking elements in the Fog Computing model are referred to as the fog nodes [8].

autonomic systems. The IBM autonomic system model consists of two main components, namely the *autonomic manager* and the *managed element*. The autonomic manager represents the control loop that manages and regulates the functionality and performance of the system under consideration (the managed element). These two components together constitute the autonomic element according to The IBM autonomic system model. Fig. 2 shows the arrangement and interactions between the involved components of this model.

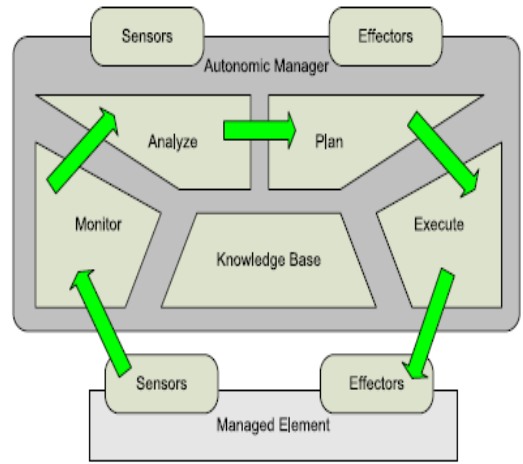

**Fig 2: IBM Autonomic element [10].**

Below is a description of these two components in the context of IoT platform.

- **Managed Element**: it represents the services provided by the physical devices that interact with each other to achieve a particular goal (business process). The system could be provided by only one service. For instance, the system goal might be monitoring the room temperature in a hotel. However, most real IoT systems consist of a number of services offered by the interaction of a set of smart devices or things. The managed system exposes some important parameters to be monitored through a set of sensors and altered via a set of actuators.
- **Autonomic Manager**: it consists of five components responsible for managing the *managed element*. They are referred to as *monitor*, *analyze*, *plan*, *execute* and *knowledge base*. The component of one autonomic manager or the feedback control loop are often distributed and not necessarily reside at the same execution environment. Moreover, most IoT applications require more than one autonomic manager to control and regulate the functionality of these applications. In fact, adopting the Fog Computing model, which is driven primarily by achieving low latency, imposes some specific organisation of the control loop components. The *autonomic manager* can also be a managed element and this explains the existence of the managerial interfaces, in the form of sensors and effectors (actuators), as depicted in Fig. 2.

*D. Distribution and Decentralization concepts*

Distribution and decentralization are two important concepts that affect the design and architecture of Fog computing based IoT applications. Description and discussion of these two concepts are presented in [11]. The distribution concept is concerned with the deployment of the software of the managed element and autonomic manager to the execution platform (hardware). A distributed autonomic system is composed of a number of software components deployed on multiple nodes connected via some network infrastructure. The other option is to deploy the autonomic system on a single node.

Decentralization here refers to a type of control in which multiple components responsible for one of the activities (monitoring for instance) of autonomic systems perform their functionality locally, but coordinated with peers. It means the monitor coordinates with other monitors, the analyser coordinates with other analysers and so on. Contrary to the decentralized coordination is the centralized one in which a single component (such as the analyser) exists to accomplish its function. The four activities of the autonomic manager are either decentralized or centralized regardless of the deployment way of the autonomic manager and managed element. In the context of IoT application adopting the fog computing approach, the deployment process is very often performed in the distributed form. This can be put down to the fact that some of the analysis and storage activities, which require powerful computation capabilities, are conducted in the cloud platform.

*E. Interaction Types in Autonomic Element*

In [11][12], the authors present a description of the various types of interactions that may occur between the managed element and the autonomic manager as well as the interactions between the different components of the autonomic manager. They classify these interactions as follows:

- Autonomic manager to managed element interaction: such an interaction occurs via the monitor component in order to perform the monitoring activity and the execute component to carry out the adaptation plans. The managed element here is the application logic which is represented by the services offered by the IoT devices. It can also be the autonomic manager itself in which case an autonomic manager is managed by another autonomic manager.
- Inter- component interaction: this interaction takes place between the different components of one autonomic manager or control loop. In a typical scenario, the monitor interacts with the analyze and the analyze interacts with the plan and the plan interacts with the execute.
- Intra-component interaction: this kind of interaction occurs between components of the same type. This kind of interactions can take two forms: the delegation and coordination. Examples include the interaction of two analyzers to coordinate the decision of issuing an adaptation request or the coordination of two executors to synchronize the adaptation or corrective actions process.

**III. PROPOSED ARCHITECTURE FOR FOG COMPUTING BASED IOT APPLICATIONS**

The proposed architecture for IoT Applications presented in this paper is built on some concepts and models discussed in the background section. The architecture is viewed as consisting of two fundamental layers: the managed element and managing element. The following subsections are dedicated to introduce the design and architecture of these two layers as well as any justifications and explanations about any design decisions made in this proposed architecture.

*A. Modelling of Managed System*

The managed system as pointed earlier represents the application logic of the system to be developed. Here are some concepts and a set of terminology we employ when modelling the system in question.

- **Domain:** The domain here is the system in question which comprises a set of tasks. Examples of domain include the healthcare, home automation, smart metering and smart building.
- **Task:** A task is a high level goal that is addressed in order to realize the overall system requirements. Each task, in turn, encompasses a set of services responsible for achieving that task. A task in a healthcare system is, for example, monitor remotely blood sugar level for a diabetic patient.
- **Service:** A service is an abstraction of a software (virtual entity) or hardware entity (physical entity or device) that plays a role in addressing the task goal. These services, later at the code generation stage, are represented as software components such as RESTful web services. A temperature sensor is an example of service. In our approach, each device or thing involved in IoT applications is treated as a service.
- **Composite:** The services of a particular task interact and coordinate with each other to address the purpose of that task. Such coordination is encapsulated in an entity called composite. A composite might consist of only one service. However, a useful composite is often composed of a number of services. Fig. 3 shows a diagrammatic view of the IoT managed system according to our approach (S refers to the service).

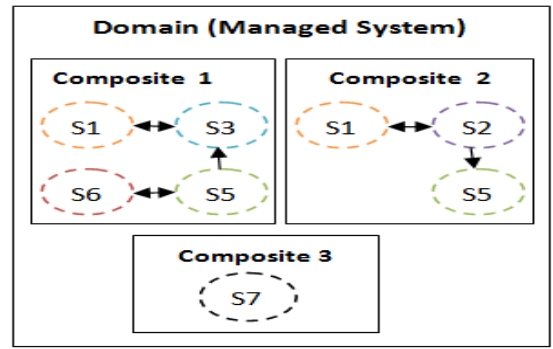

**Fig 3: A Proposed Managed System Model.**

### B. Modelling of Managing System

The managing system represents the control loop that controls and regulates the functionality of the managed system. The four components, in addition to the knowledge base component, of the control loop which are responsible for the monitoring, analysis, plan and execute activities are modelled and hosted on a set of fog nodes located proximate to the physical devices or things that provide the services of the system in question (managed system). We here discuss the layout and arrangement of the control loop components over the fog computing platform as well as the cloud. The proposed architecture for the control loop is driven by the following requirements:

- *The control and regulation of the functionality of IoT applications must be conducted on a timely manner.*
- *Powerful computing, analysis and storage capabilities should be provided to meet the requirements of large scale and complex IoT applications.*
- *The support for the splitting up of the local control loop into a set of smaller control loops with each one responsible for controlling and regulating a particular area in the same application in a wide deployment area.*
- *The support and provision of the coordination between the local control loops to regulate the functionality of the managed system in a decentralized mode.*
- *The delegation of one or more activities of the control loop to one or more local control loops and regulate the managed system in a centralized mode.*

To meet the above stated requirements, we have deployed a local control loop on fog nodes nearby the device layer where the services of the managed system are provided. We offer this control function as a **MAPEaaService** in the fog computing platform. We also offer the same service on the cloud platform to cater for the need of powerful computation and storage capabilities when developing large and big data generating applications. The control loop at the cloud contains only, in addition to the knowledge component, the analysis and planning activities. Thus, we refer to this service as a **APaaService.** We offer two modes of control: *centralized* and *decentralized*. In the *centralized mode*, a central control loop is deployed either on the fog computing or cloud platform (depends on the application scale) to regulate the operating of the different control loops that reside at the same level. We draw the relationship between the central and local control loops using the master-slave model. The local control loop is in charge of controlling the functionality of a sub system, where monitoring and keeping values of interesting parameters related to this sub system is taken place. In contrast, the central control loop regulates the working of the whole system. This usually involves monitoring and keeping values of interesting parameters at a desirable range related to the whole system. Such interesting parameters represent the system state which can be formed by combining a set of parameters from the sub systems. These parameters can be of the same type as the case where the central control loop monitors and controls the energy consumption of a set of offices in a building or a set of buildings in a city. Also, the system state can be composed of parameters of different types. A typical example of this case is the monitoring of a patient condition in the healthcare application where his/her condition is diagnosed by a number of different readings such as the temperature, blood pressure, blood sugar, etc. The arrangement of this mode is depicted in Fig. 4.

In the de*centralized mode,* a set of control loops of the same level is coordinated to accomplish the four activities (monitoring, analysis, planning and execution). For instance, the execute components of each control loop communicate and coordinate to carry out the corrective actions in the absence of a central controller. Figure 5 depicts the organization of this mode of control and regulation. Self organising systems are a popular example of systems operating and functioning in the decentralised mode.

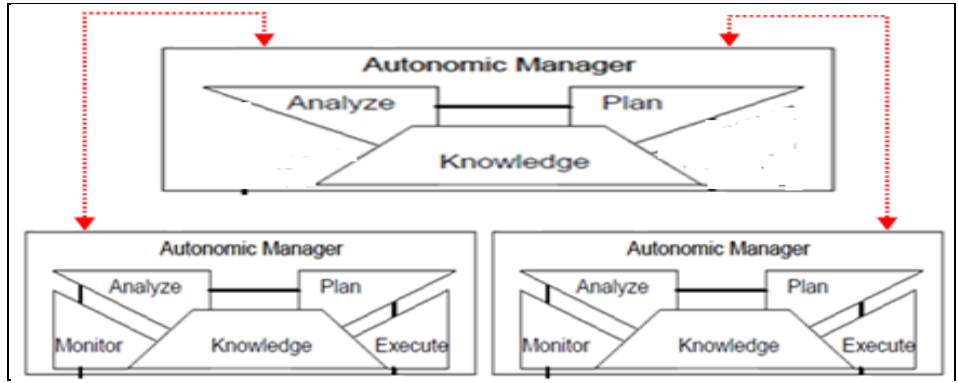

**Fig 4: A Centralized mode of control loop.**

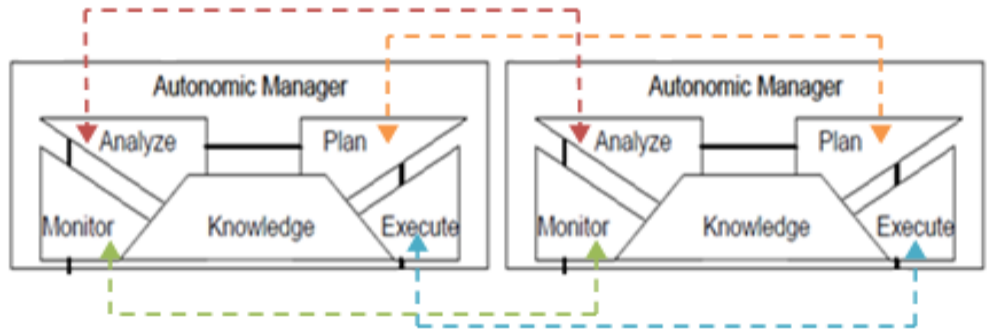

**Fig 5: A Decentralized mode of control loop.**

## IV. AN ILLUSTRATIVE CASE STUDY: SMART BUILDING

To demonstrate the feasibility of the proposed architecture, we introduce here the case study of the smart building. The smart building here consists of a number of smart offices. Each smart office should address and meet the following requirements:

- *Security measures should be provided in terms of who is authorized to get in.*
- *Room temperature should be kept at a reasonable level.*
- *The power should be consumed reasonably and efficiently inside the office. For example, the lights should be automatically turned off when it is sunny and the office window is open.*
- *The office should be ventilated occasionally and when needed.*
- *Provide a facility to measure the consumed energy.*

The following devices or things are needed in this smart office based on the above requirements:
- A smart door
- A smart window
- A smart heater
- A smart energy meter
- A smart lamp
- A smart clock

### A. Scenarios of interactions

There will be a lot of interactions and coordination between the involved devices to address both the individual goal of each device as well as the overall goal of the system (the smart office). The interaction and coordination between the different devices or things may take different forms at different occasions. These forms of interaction will be primarily driven by the requirements and goals outlined earlier. One action of one device could be triggered by a change on another device. To keep the room temperature at a certain level, for instance, the smart heater will probably trigger the office window to perform a certain action (e.g. open) or the other way around. This also addresses the goal of consuming the power efficiently (the heater is switched off and the window is either open or closed). Another possible scenario might happen when the office owner forgets to, for example, turn the lights off upon leaving. In this case, the light switch is triggered by the information coming from both the smart clock and smart door. Upon locking the door, a signal is sent out to the smart clock to start timing. Once the specified time has passed, the lights must be turned off.

### B. *Control loop architecture for Smart building*

As stated earlier, the smart building is composed of a number of smart offices where each office is controlled by one separate control loop. The managed system here represents the services provided by the devices located at each office. The whole control loop process is driven by the parameter to be monitored and regulated. In this case study, we assume that the main concern of the smart building is to consume the energy in an efficient manner which requires each office to turn on the heater only when needed as described in the smart office requirements. Thus, the parameter of great concern here is the energy meter reading at each office. These readings collectively constitute the system state of the smart building application. Fig. 6 shows the architecture of the control loops for the smart building system where a centralized mode of control is employed.

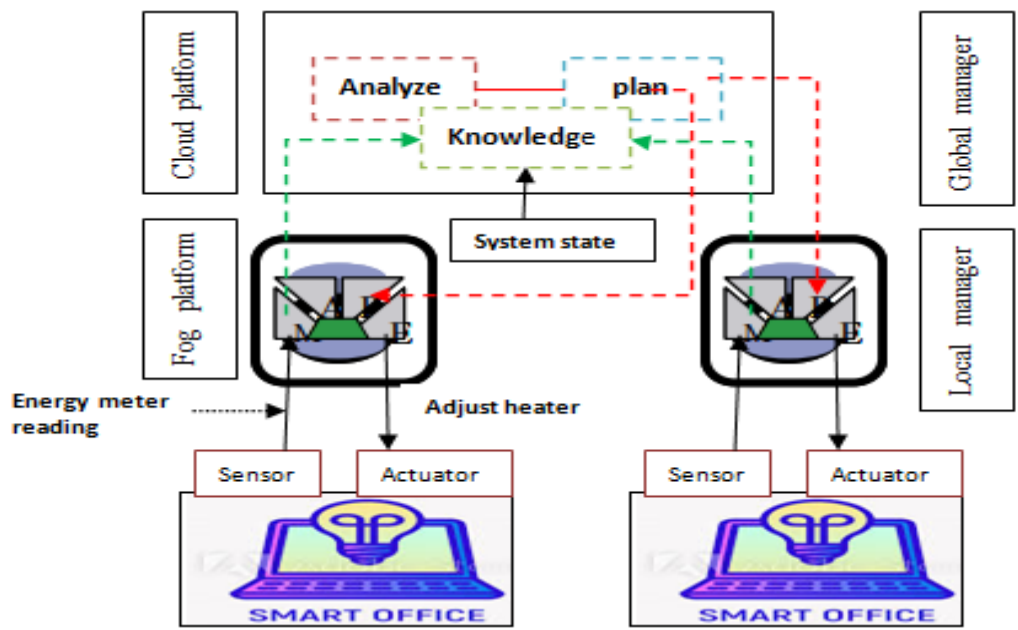

**Fig 6 Proposed architecture for control loops for smart building application.**

## V. RELATED WORK

Despite its recent emergence, a great deal of research papers and studies have been published in the area of Fog Computing. An early study was conducted by Bonomi et al [13] in which an architecture of Fog Computing platform was proposed. The authors in this research defined and specified a number of characteristics which made the Fog Computing worth considering and looked a promising solution. They also highlighted the applications and services that could highly benefit from the Fog Computing which include Connected Vehicle, Smart Grid , Smart Cities, and, in general, Wireless Sensors and Actuators Networks (WSANs). Another work by [14] proposed an architecture for the Fog Computing which was inspired by the human nervous system. In such an architecture, the cloud data centre represents the brain nerve centre, the Fog Computing data centre represents the spinal nerve centre and smart devices represents peripheral nerve centres. Aazam et al [15] proposed a six layer architecture for the Fog computing platform. These layers include, from bottom to top, the physical layer, monitoring later, pre-processing layer, temporary storage layer, security layer and pre-processed data uploading layer. In a similar work by Dastjerdi et al [16], the Fog platform is architecting using five layers: the application layer, management layer (monitoring, security, etc), cloud service management layer, network layer and physical layer. Another layered architecture was presented by Arkian et al [17] in which the Fog Computing platform is composed of four layers, namely the data generator layer, Cloud computing layer, Fog computing layer and data consumer layer. To provide an open reference architecture for Fog Computing, the OpenFog Consortium was founded in 2015 by members from ARM, Cisco, Dell, Intel, Microsoft, and Princeton University. Later, this consortium released the OpenFog reference architecture [8]. To the best of our knowledge, none of the approaches proposed so far has tackled the subject of developing IoT applications at the Fog Computing using the control loops from the area of autonomic systems. Our approach is different in that it is based on the IBM architecture blueprint in which the fundamental components of the control loop (monitoring, analysis, planning and execution) are modelled as first class entities.

## VI. CONCLUSIONS AND FUTURE WORK

In this paper, we have proposed an architecture for IoT applications hosted on the Fog computing platform. We discussed the high level architecture of Fog computing and its benefits for the design and development of IoT applications. Since IoT applications are highly dynamic in nature and involve a great deal of monitoring and analysis activities, we have found it helpful to engineer these applications by employing some concepts and models from the self adaptive and autonomic systems. Our proposed architecture was thus based on the IBM architecture blueprint for autonomic systems. We also showed the impact of hosting IoT applications on the fog computing platform on the arrangement and distribution of the control loop components over a number of nodes. In particular, we deployed a local control loop on fog nodes nearby the device layer where the services of the managed system are provided. We offer this control function as a MAPEaaService in the fog computing platform. We also offer the same service on the cloud platform to cater for the need of powerful computation and storage capabilities when developing large and big data generating applications in the form of APaaService. We offer two modes of control: centralized and decentralized in our proposed architecture. For future work, the following issues need to be addressed:

- A more detailed and different use case is needed to evaluate and illustrate the feasibility of the proposed architecture.
- A more detailed design for each activity (monitoring for example) of the control loop in our proposed architecture; each activity contains a number of involved components and interactions and it is complex enough to be treated separately.
- The investigation of the impact of the application type (healthcare for instance) on the control mode (centralized and decentralized) of our proposed architecture.